\documentclass[aps,prl,onecolumn,superscriptaddress,showpacs,preprintnumbers,amsmath,amssymb]{revtex4}

\usepackage{graphicx} % Include figure files
\usepackage{dcolumn}  % Align table columns on decimal point
\usepackage{subfigure}

\newcommand{\ee}{\ensuremath{e^+ e^-}}
\newcommand{\cc}{\ensuremath{c\bar{c}}}
\newcommand{\bb}{\ensuremath{B\overline{B}}}
\newcommand{\tg}{\ensuremath{{\text{tag}}}}
\newcommand{\ccr}{\ensuremath{(\cc)_{\mathrm{res}}}}

\newcommand{\RM}{\ensuremath{M_{\mathrm{recoil}}}}
\newcommand{\nch}{\ensuremath{N_{\mathrm{ch}}}}

\newcommand{\gevc}{\ensuremath{\, {\mathrm{GeV}/c^2}}}
\newcommand{\gevp}{\ensuremath{\, {\mathrm{GeV}/c}}}
\newcommand{\mevc}{\ensuremath{\, {\mathrm{MeV}/c^2}}}
\newcommand{\gev}{\ensuremath{\, {\mathrm{GeV}}}}
\newcommand{\mev}{\ensuremath{\, {\mathrm{MeV}}}}

\newcommand{\ifb}{\ensuremath{\, {\mathrm{fb^{-1}}}}}
\newcommand{\pb}{\ensuremath{\, {\mathrm{pb}}}}

\newcommand{\BR}{\ensuremath{\cal{B}}}
\newcommand{\el}{\ensuremath{\ell^+ \ell^-}}
\newcommand{\mel}{\ensuremath{M_{\ell^+ \ell^-}}}

\newcommand{\thel}{\ensuremath{\theta_{\rm{hel}}}}
\newcommand{\tprod}{\ensuremath{\theta_{\rm{prod}}}}
\newcommand{\chel}{\ensuremath{\left|\cos\theta_{\rm{hel}}\right|}}
\newcommand{\cprod}{\ensuremath{\left|\cos\theta_{\rm{prod}}\right|}}
\newcommand{\ahel}{\ensuremath{\alpha_{\rm{hel}}}}
\newcommand{\aprod}{\ensuremath{\alpha_{\rm{prod}}}}

\newcommand{\ups}{\ensuremath{\Upsilon(4S)}}
\newcommand{\sig}{\ensuremath{\sigma}}
\newcommand{\hc} {\ensuremath{H_c}}
\newcommand{\jp} {\ensuremath{J/\psi}}
\newcommand{\pp} {\ensuremath{\psi^\prime}}
\newcommand{\cha} {\ensuremath{\chi_{c1(2)}}}
\newcommand{\cho} {\ensuremath{\chi_{c1}}}
\newcommand{\cht} {\ensuremath{\chi_{c2}}}

\newcommand{\pjp} {\mbox{\ensuremath{p^{*}_{J/\psi}}}}

\newcommand{\Do}{\mbox{\ensuremath{D^0}}}
\newcommand{\Ds}{\mbox{\ensuremath{D^+_s}}}
\newcommand{\Dp}{\mbox{\ensuremath{D^+}}}
\newcommand{\Lc}{\mbox{\ensuremath{\Lambda_c^+}}}
\newcommand{\La}{\mbox{\ensuremath{\Lambda^0}}}
\newcommand{\ks}{\mbox{\ensuremath{K^0_S}}}

\newcommand{\rescc}{\mbox{\ensuremath{0.74 \pm 0.08}}}
\newcommand{\resncc}{\mbox{\ensuremath{0.43 \pm 0.09}}}
\newcommand{\resx}{\mbox{\ensuremath{1.17 \pm  0.02}}}

\newcommand{\syscc}{\mbox{\ensuremath{{\,}^{+0.09}_{-0.08}}}}
\newcommand{\sysncc}{\mbox{\ensuremath{{}\pm 0.09}}}
\newcommand{\sysx}{\mbox{\ensuremath{{} \pm 0.07}}}

\newcommand{\jpcc}{\ensuremath{ \jp \,{c\bar{c}}}}
\newcommand{\jpncc}{\mbox{\ensuremath{ \jp \, X_{\text{non-}c\bar{c}}}}}
\newcommand{\jpx}{\mbox{\ensuremath{ \jp \, X}}}

\newcommand{\jph}{\mbox{\ensuremath{ \jp \, H_c\,X}}}

\newcommand{\eejpgg}{\ensuremath{ \ee \to \jp \, gg}}
\newcommand{\eejpg}{\ensuremath{ \ee \to \jp \, g}}
\newcommand{\eejph}{\mbox{\ensuremath{ \ee \to \jp \, \hc \, X}}}
\newcommand{\eejpcc}{\ensuremath{\ee \to \jp \,{c\bar{c}}}}
\newcommand{\eejpncc}{\mbox{\ensuremath{\ee \to \jpncc}}}
\newcommand{\eejpx}{\mbox{\ensuremath{\ee \to \jp \, X}}}

\begin{document}

\preprint{\vbox{ \hbox{   }
                 \hbox{Belle Preprint 2009-2}
                 \hbox{KEK Preprint 2009-5}
}}

\vspace*{-3\baselineskip}
\resizebox{!}{2.6cm}{\includegraphics{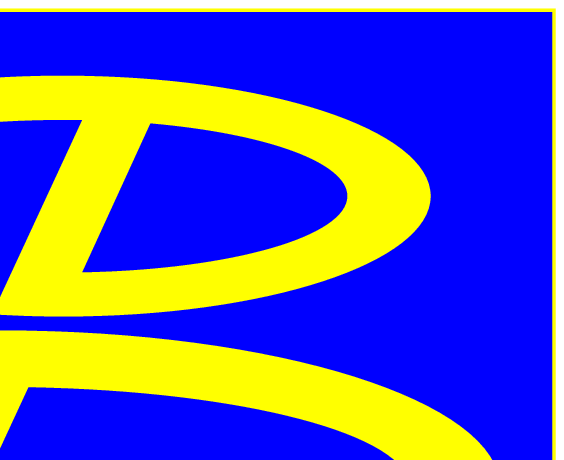}}

\title{ \quad\\[0.5cm] \Large Measurement of the \eejpcc\ cross
  section at $\mathbf{\sqrt{s} \approx 10.6}\,$GeV}

\pacs{13.66.Bc,12.38.Bx,14.40.Gx}

%%% Use \input{author} to insert this material into your latex file.
%%%%% Force institutions to appear in alphabetical order when typeset.
\affiliation{Budker Institute of Nuclear Physics, Novosibirsk}
\affiliation{Chiba University, Chiba}
\affiliation{University of Cincinnati, Cincinnati, Ohio 45221}
\affiliation{T. Ko\'{s}ciuszko Cracow University of Technology, Krakow}
\affiliation{Department of Physics, Fu Jen Catholic University, Taipei}
\affiliation{Justus-Liebig-Universit\"at Gie\ss{}en, Gie\ss{}en}
%%%\affiliation{The Graduate University for Advanced Studies, Hayama}
%%%\affiliation{Gyeongsang National University, Chinju}
\affiliation{Hanyang University, Seoul}
\affiliation{University of Hawaii, Honolulu, Hawaii 96822}
\affiliation{High Energy Accelerator Research Organization (KEK), Tsukuba}
\affiliation{Hiroshima Institute of Technology, Hiroshima}
%%%\affiliation{University of Illinois at Urbana-Champaign, Urbana, Illinois 61801}
\affiliation{Institute of High Energy Physics, Chinese Academy of Sciences, Beijing}
\affiliation{Institute of High Energy Physics, Vienna}
\affiliation{Institute of High Energy Physics, Protvino}
%%%\affiliation{INFN - Sezione di Torino, Torino}
\affiliation{Institute for Theoretical and Experimental Physics, Moscow}
\affiliation{J. Stefan Institute, Ljubljana}
\affiliation{Kanagawa University, Yokohama}
\affiliation{Korea University, Seoul}
%%%\affiliation{Kyoto University, Kyoto}
\affiliation{Kyungpook National University, Taegu}
\affiliation{\'Ecole Polytechnique F\'ed\'erale de Lausanne (EPFL), Lausanne}
\affiliation{Faculty of Mathematics and Physics, University of Ljubljana, Ljubljana}
\affiliation{University of Maribor, Maribor}
\affiliation{University of Melbourne, School of Physics, Victoria 3010}
\affiliation{Nagoya University, Nagoya}
\affiliation{Nara Women's University, Nara}
\affiliation{National Central University, Chung-li}
\affiliation{National United University, Miao Li}
\affiliation{Department of Physics, National Taiwan University, Taipei}
\affiliation{H. Niewodniczanski Institute of Nuclear Physics, Krakow}
\affiliation{Nippon Dental University, Niigata}
\affiliation{Niigata University, Niigata}
\affiliation{University of Nova Gorica, Nova Gorica}
\affiliation{Novosibirsk State University, Novosibirsk}
\affiliation{Osaka City University, Osaka}
%%%\affiliation{Osaka University, Osaka}
\affiliation{Panjab University, Chandigarh}
%%%\affiliation{Peking University, Beijing}
%%%\affiliation{Princeton University, Princeton, New Jersey 08544}
%%%\affiliation{RIKEN BNL Research Center, Upton, New York 11973}
\affiliation{Saga University, Saga}
\affiliation{University of Science and Technology of China, Hefei}
\affiliation{Seoul National University, Seoul}
%%%\affiliation{Shinshu University, Nagano}
\affiliation{Sungkyunkwan University, Suwon}
\affiliation{University of Sydney, Sydney, New South Wales}
\affiliation{Tata Institute of Fundamental Research, Mumbai}
\affiliation{Toho University, Funabashi}
\affiliation{Tohoku Gakuin University, Tagajo}
\affiliation{Tohoku University, Sendai}
\affiliation{Department of Physics, University of Tokyo, Tokyo}
\affiliation{Tokyo Institute of Technology, Tokyo}
\affiliation{Tokyo Metropolitan University, Tokyo}
\affiliation{Tokyo University of Agriculture and Technology, Tokyo}
%%%\affiliation{Toyama National College of Maritime Technology, Toyama}
\affiliation{IPNAS, Virginia Polytechnic Institute and State University, Blacksburg, Virginia 24061}
\affiliation{Yonsei University, Seoul}
   \author{P.~Pakhlov}\affiliation{Institute for Theoretical and Experimental Physics, Moscow} % ITEP
% \author{I.~Adachi}\affiliation{High Energy Accelerator Research Organization (KEK), Tsukuba} % KEK
   \author{H.~Aihara}\affiliation{Department of Physics, University of Tokyo, Tokyo} % Tokyo
   \author{K.~Arinstein}\affiliation{Budker Institute of Nuclear Physics, Novosibirsk}\affiliation{Novosibirsk State University, Novosibirsk} % BINP
% \author{T.~Aso}\affiliation{Toyama National College of Maritime Technology, Toyama} % Toyama
% \author{V.~Aulchenko}\affiliation{Budker Institute of Nuclear Physics, Novosibirsk}\affiliation{Novosibirsk State University, Novosibirsk} % BINP
   \author{T.~Aushev}\affiliation{\'Ecole Polytechnique F\'ed\'erale de Lausanne (EPFL), Lausanne}\affiliation{Institute for Theoretical and Experimental Physics, Moscow} % ITEP
% \author{T.~Aziz}\affiliation{Tata Institute of Fundamental Research, Mumbai} % Tata
% \author{S.~Bahinipati}\affiliation{University of Cincinnati, Cincinnati, Ohio 45221} % Cincinnati
  \author{A.~M.~Bakich}\affiliation{University of Sydney, Sydney, New South Wales} % Sydney
   \author{V.~Balagura}\affiliation{Institute for Theoretical and Experimental Physics, Moscow} % ITEP
% \author{Y.~Ban}\affiliation{Peking University, Beijing} % Peking
   \author{E.~Barberio}\affiliation{University of Melbourne, School of Physics, Victoria 3010} % Melbourne
   \author{A.~Bay}\affiliation{\'Ecole Polytechnique F\'ed\'erale de Lausanne (EPFL), Lausanne} % Lausanne
% \author{I.~Bedny}\affiliation{Budker Institute of Nuclear Physics, Novosibirsk}\affiliation{Novosibirsk State University, Novosibirsk} % BINP
   \author{K.~Belous}\affiliation{Institute of High Energy Physics, Protvino} % Protvino
   \author{V.~Bhardwaj}\affiliation{Panjab University, Chandigarh} % Panjab
% \author{S.~Blyth}\affiliation{National United University, Miao Li} % NUU
% ...
% BINP
\collaboration{The Belle Collaboration}

\begin{abstract}
We present a new measurement of the \eejpcc\ cross section where the
\cc\ pair can fragment either into charmed hadrons or a charmonium
state.  In the former case the \jp\ and a charmed hadron are
reconstructed, while the latter process is measured using the recoil
mass technique, which allows the identification of two-body final
states without reconstruction of one of the charmonia.
%We present a new measurement of the \eejpcc\ cross section where the
%\cc\ pair can fragment either into a charmonium state or charmed
%hadrons. The former process is measured using the recoil mass
%technique, which allows the identification of two-body final states
%without reconstruction of one of the charmonia. The latter is measured
%using the reconstruction of the \jp\ and a charmed hadron. 
The measured \eejpcc\ cross section is $(\rescc \syscc)\,$pb, and the
\eejpncc\ cross section is $(\resncc \sysncc)\,$pb. We note that the
measured cross sections are obtained from a data sample with the
multiplicity of charged tracks in the event larger than four;
corrections for the effect of this requirement are not performed as
this cannot be done in a model-independent way. The analysis is based
on a data sample with an integrated luminosity of $673\ifb$ recorded
near the \ups\ resonance with the Belle detector at the KEKB
\ee\ asymmetric-energy collider.
\end{abstract}

\maketitle
\setcounter{footnote}{0} 

Prompt charmonium production in \ee\ annihilation is important for
studying the interplay between perturbative QCD and non-perturbative
effects. The production rate and kinematic characteristics of
\jp\ mesons in \ee\ annihilation are poorly described by theory, and
even the production mechanisms are not understood. An effective field
theory, non-relativistic QCD (NRQCD), predicts that prompt
\jp\ production at $\sqrt{s}\!\approx \!  10.6\,\gev$ is dominated by
$\ee \to \jp \, gg$ with a $1\pb$ cross section~\cite{nrqcd2a}; the
$\ee \to \jp \, g$ contribution, which may be of the same order, is
uncertain due to poorly-constrained color-octet matrix
elements~\cite{nrqcd2d}.  The \eejpcc\ cross section is predicted to
be $\sim \!  0.05-0.1\pb$~\cite{lik}, only $\sim \!10\%$ of that for
$\jp \, gg$~\cite{lik2}.  (The estimate of the ratio is more precise,
as QCD uncertainties partially cancel.)  By contrast, Belle observed
the ratio of the \jpcc\ and inclusive \jp\ production cross sections
to be $0.59^{+0.15}_{-0.13} \pm 0.12$~\cite{2cc}, and thus found
$\sigma (\eejpcc)/ \sigma(\ee \to \jp \, gg) \gtrsim 1$. Such a large
value cannot be explained within the NRQCD framework, however some
alternative approaches (see {\em {e.g.}} Ref.~\cite{kai}) can
accommodate it.

In this report we present a new measurement of the \eejpcc\ cross
section. This process can be experimentally tagged by the presence of
another charmed particle (either charmonium or charmed hadrons) in the
event in addition to the reconstructed \jp. The technique used in this
analysis allows the model dependence of the result to be removed,
reducing the systematic uncertainties. Production of the \jp\ via
mechanisms other than \eejpcc\ is also studied. The \jp\ momentum
spectrum, and helicity and production angle distributions, are
measured for both \eejpcc\ and \jpncc\ processes.  The analysis is
performed using data recorded at the \ups\ and in the continuum
$60\mev$ below the resonance, corresponding to integrated luminosities
of $605\ifb$ and $68\ifb$, respectively. The data are collected with
the Belle detector~\cite{Belle} at the KEKB asymmetric-energy
\ee\ collider~\cite{KEKB}.

The Belle detector is a large-solid-angle magnetic spectrometer that
consists of a silicon vertex detector (SVD), a 50-layer central drift
chamber (CDC), an array of aerogel threshold Cherenkov counters (ACC),
a barrel-like arrangement of time-of-flight scintillation counters
(TOF), and an electromagnetic calorimeter (ECL) comprised of CsI(Tl)
crystals located inside a superconducting solenoid coil that provides
a $1.5\,$T magnetic field. An iron flux-return located outside the
coil is instrumented to detect $K_L^0$ mesons and to identify muons
(KLM). Two inner detector configurations were used. A $2.0\,$cm
beampipe and a 3-layer silicon vertex detector were used for the first
sample of $\sim \! 156\ifb$, while a $1.5\,$cm beampipe, a 4-layer
silicon detector and a small-cell inner drift chamber were used to
record the remaining data sample.

We use a selection procedure similar to that described in
Ref.~\cite{2cc}. All charged tracks are required to be consistent with
originating from the interaction point (IP); we impose the
requirements $dr\!<\!2 \, {\mathrm{cm}}$ and
$|dz|\!<\!4\,{\mathrm{cm}}$, where $dr$ and $dz$ are the impact
parameters perpendicular to and along the beam direction with respect
to the IP.  Particle identification requirements are based on CDC, ACC
and TOF information~\cite{nim}. Charged kaon and proton candidates are
required to be positively identified: the identification efficiencies
typically exceed 90\%, while misidentification probabilities are less
than 10\%. No identification requirements are applied for pion
candidates, as the pion multiplicity is much higher than those of
other hadrons. \ks\ (\La) candidates are reconstructed by combining
$\pi^+ \pi^-$ ($p \pi^-$) pairs with an invariant mass within
$10\mevc$ of the nominal \ks\ (\La) mass. We require the distance
between the tracks at the \ks\ (\La) vertex to be less than
$1\,\mathrm{cm}$, the transverse flight distance from the IP to be
greater than $1\,\mathrm{mm}$ and the angle between the \ks\ (\La)
momentum direction and its decay path to be smaller than
$0.1\,\mathrm{rad}$. Photons are reconstructed in the ECL as showers
with energies more than $50 \mev$ that are not associated with charged
tracks.

\jp\ candidates are reconstructed via the $\jp \!  \to \!  \el$
($\ell=e, ~\mu$) decay channel. Two positively identified lepton
candidates are required to form a common vertex that is less than
$1\,\mathrm{mm}$ from the IP in the plane perpendicular to the beam
axis ($\approx\! 98\%$ efficiency). A partial correction for final
state radiation and bremsstrahlung energy loss is performed by
including the four-momentum of every photon detected within a
$50\,\mathrm{mrad}$ cone around the electron and positron direction in
the \ee\ invariant mass calculation. The \jp\ signal region is defined
by the mass window $\left|M_{\ell^{+} \ell^{-}} - m_{J/\psi}\right| \!
< \! 30 \mevc$ ($\approx \! 2.5\, \sigma$). A mass-constrained fit is
then performed for the signal window candidates, to improve the
center-of-mass (CM) momentum \pjp\ resolution. QED processes are
suppressed by requiring the total charged multiplicity (\nch) in the
event to be greater than 4. In the \ups\ data \jp\ mesons from
$B\overline{B}$ events are removed by requiring $\pjp\!>\!2.0\gevp$;
no requirement on \pjp\ is applied in the off-resonance data sample.

We also reconstruct charmonia decaying to \jp.  \pp\ candidates are
reconstructed via the decay to $\jp \,\pi^+ \pi^-$, with the
\pp\ signal window defined by $|M_{J/\psi\, \pi^+ \pi^-} -
m_{\psi^\prime}| \! <\! 10\,\mevc$ ($\approx \!  3\, \sigma$).
$\chi_{c1}$ and $\chi_{c2}$ candidates are reconstructed using the
$\jp\,\gamma$ mode; signal windows of $\pm 20\mevc$ are chosen around
the corresponding nominal masses ($\approx \! 2.5\, \sigma$). In
addition we require $\cos\theta_\gamma \! < \!0$, where
$\theta_\gamma$ is defined as the angle between the photon momentum
and the CM system, seen from the \cha\ rest frame. This requirement
suppresses the large combinatorial background due to low energy
photons by more than an order of magnitude, while retaining 50\% of
the signal, independent of the \cha\ polarization.

We use only charged final states for charmed hadron reconstruction to
avoid correlated multiple candidates. Candidate \Do\ mesons are
reconstructed in the $K^- \pi^+$, $K^+ K^-$, $\ks \pi^+ \pi^-$ and
$K^- \pi^- \pi^+ \pi^+$ decay modes~\cite{cc}. We reconstruct
\Dp\ mesons using $K^- \pi^+ \pi^+$, $K^- K^+ \pi^+$, $\ks \pi^+$ and
$\ks \pi^+ \pi^+ \pi^-$ decays; for \Ds\ meson reconstruction we use
the $K^- K^+ \pi^+$ and $\ks \pi^+$, and finally \Lc\ baryons are
reconstructed via $p K^- \pi^+$, $p \ks$ and $\La \pi^+$. A $\pm
15\mevc$ mass window ($\approx\!  2.5\,\sigma$) is used throughout,
except for the $\Do \to K^- \pi^- \pi^+ \pi^+$ and $\Dp \to \ks \pi^+
\pi^+ \pi^-$ modes where the resolution is better, and the
combinatorial background higher: in these cases a $\pm 10\mevc$ window
is chosen ($\approx\!  2.3\,\sigma$).  To study the contribution of
combinatorial background under the various charmed hadron peaks, we
use sidebands selected from a mass window four times as large.

We generate large Monte Carlo (MC) samples of double charmonium
production and of the process \eejpcc\ with fragmentation to open
charm. We also generate a sample of $\ee \to \jp \,q\bar{q}$ events
for the study of the \eejpncc\ process. In the MC samples the
\jp\ kinematical characteristics (momentum spectrum and angular
distributions) are tuned to those measured in the data. As the
measured distributions are extracted from the data using the MC
simulation, the tuning procedure is repeated until the difference
between successive iterations becomes negligibly small.

To measure the contribution of \cc\ resonances to the \eejpcc\ cross
section, we reconstruct all double charmonium final states that can
result in the presence of a \jp\ in the event: $\jp \ccr$, $\pp \ccr$,
and $\cha \ccr$, where \ccr\ is one of the charmonium states below
open-charm threshold. If a charmonium state lies above the open-charm
threshold~\cite{x4160}, we assume it will decay predominantly to
charmed hadrons; production of a \jp\ together with charmed hadrons is
treated separately below.  The process $\ee \to Y \ccr$, where $Y$ is
one of the states odd under charge conjugation, recently observed in
initial state radiation (ISR) studies~\cite{y4260}, can produce
\jp\ from $Y$ decays. However, we are unable to measure this
contribution because of the large intrinsic width of the $Y$ states,
and ignore it. Following the method described in~\cite{2cc, 2cc2} we
first reconstruct a $(\cc)_{\tg}=\jp$, \pp, or \cha\ meson to tag the
process, and then form the recoil mass
\begin{equation}
\RM((\cc)_{\tg}) = \sqrt{(E_{\rm CM}-E_{\tg}^{*})^2-p_{\tg}^{*~2}},
\end{equation} 
where $E^*_{\tg}$ and $p_{\tg}^*$ are the CM energy and momentum of
the reconstructed charmonium, and $E_{\rm CM}$ is the CM energy. The
$\RM((\cc)_{\tg})$ spectra for the data are presented in
Fig.~\ref{cc2}. We assume that only
charmonium states with a charge conjugation eigenvalue opposite to that of
$(\cc)_{\tg}$ can appear; two virtual photon annihilation, which can
produce a pair of charmonium states with the same eigenvalue, was
not observed in Ref.~\cite{2cc2}, and is expected to be small.
%Due to charge conjugation symmetry, only $\eta_c$, $\chi_{c0(1,2)}$
%and $\eta_c'$ can be produced recoiling against \jp\ and \pp, while
%only \jp, $h_c$ and \pp\ can be observed in the $\RM(\cha)$ spectra.

We fit the four $\RM((\cc)_{\tg})$ spectra simultaneously to fix the
$\psi^{(\prime)}\,\cha$ contributions, which are poorly resolved in
the $\RM(\psi^{(\prime)})$ spectra. The ratios of the $\psi^{(\prime)}
\,\cha$ signal contributions to the $\RM(\psi^{(\prime)})$ and
$\RM(\cha)$ spectra are fixed according to the MC study. The signal
line shapes for all the double charmonium final states are obtained
from MC simulation, with ISR included, and the background is
parameterized by a linear function (a second order polynomial function
in the $\RM(\jp)$ case). Only the region below the open-charm
threshold ($\RM \!< \!3.7\gevc$) is included in the fit. The fitting
function for the $\RM(\jp)$ spectrum also includes the expected
contribution from the ISR process $\ee \to \pp \,\gamma$, which is
poorly described by the polynomial function; its shape and
normalization are fixed from the MC simulation. This process was
studied for our paper~\cite{y4260}, and the measured width
$\Gamma_{ee}(\pp)$ was found to be in good agreement with the PDG
value~\cite{pdg}.
\begin{figure}[htb]
\hspace*{-0.025\textwidth} \includegraphics[width=0.69\textwidth]
        {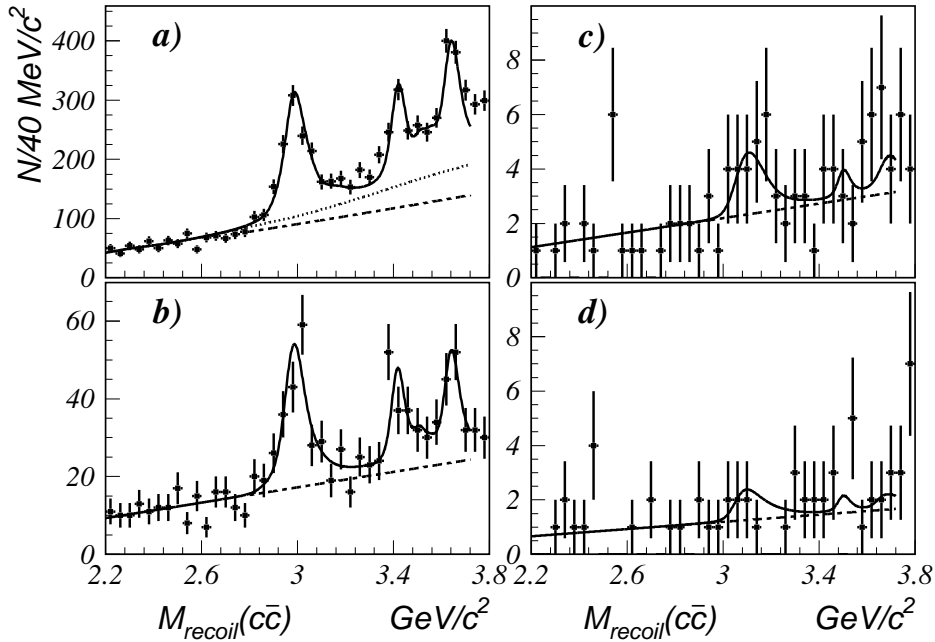}
\caption{The mass of the system recoiling against the reconstructed a)
  \jp, b) \pp, c) \cho\ and d) \cht. The curves show the fit results
  described in the text.}
\label{cc2}
\end{figure}

\begin{table}[htb]
\caption{$\ee\to (\cc)_{\tg} \ccr$ signal yields (significances)
	from a simultaneous fit to $\RM((\cc)_{\tg})$ spectra. }
\label{tab:rmx}
\begin{center}
\begin{ruledtabular}
\begin{tabular}{l|cccc}
% $\ccr$ & $(\cc)_{\tg}\!=\!\jp$ & $(\cc)_{\tg}\!=\!\pp$ &
% $(\cc)_{\tg}\!=\!\cho$ & $(\cc)_{\tg}\!=\!\cht$ \\ \hline \hline
$\ccr$	& \multicolumn{4}{c}{$(\cc)_{\tg}$:}	\\
	& $\jp$ & $\pp$ & $\cho$ & $\cht$	\\ \hline

$\eta_c$ & $1032 \pm 62\,(19)$ & $161 \pm 22\,(8.2)$ & --- & --- \\

\jp & --- & --- & $16 \pm 5\,(3.2)$ & $9 \pm 4\,(2.1)$ \\

$\chi_{c0}$ & $525 \pm 54\,(9.6)$ & $75 \pm 19\,(4.3)$ & --- & --- \\

$\chi_{c1}$ & $119 \pm 39\,(3.2)$ & $12 \pm 12$ & --- & --- \\

$h_{c}$ & --- & --- & $4 \pm 6$ & $1 \pm 5$ \\

$\chi_{c2}$ & $99 \pm 43\,(2.1)$ & $7 \pm 16$ & --- & --- \\

$\eta_c'$ & $679 \pm 63\,(10)$ & $81 \pm 19\,(4.5)$ & --- & --- \\

\pp & --- & --- & $6 \pm 6$ & $2 \pm 5$ 
\end{tabular}
\end{ruledtabular}
\end{center}
\end{table}

The fit results are shown in Fig.~\ref{cc2} by solid curves; the
background function and the $\ee \to \pp \, \gamma$ reflection are
shown with dashed and dotted curves, respectively. The signal yields
and significances for all the studied double charmonium processes are
listed in Table~\ref{tab:rmx}. The statistical significance of each
process is determined from $-2\ln(\mathcal{L}_0 /
\mathcal{L}_{\text{max}})$, where $\mathcal{L}_{\text{max}}$ is the
maximum likelihood returned by the fit, and $\mathcal{L}_0$ is the
likelihood with the corresponding contribution set to zero.  The
results for $\jp \,\ccr$ and $\pp \,\ccr$ are in good agreement with
our previous measurements~\cite{2cc,2cc2}. There is also evidence for
$\jp \, \cho$ production at the $3.2\sigma$ level (statistical only).

Next, we study associated production of a \jp\ with charmed
hadrons. In the previous paper~\cite{2cc} we determined the
\jpcc\ cross section from measurements of the production rate of a
\jp\ with associated \Do\ and $D^{*+}$ mesons using Lund~\cite{lund}
model predictions for probabilities of fragmentation $\cc \to
\Do(D^{*+})$. Moreover, to suppress combinatorial background from
\bb\ events, we applied additional kinematical criteria; the
efficiency of these criteria also contributed to the model dependence
of the result. To eliminate the model dependence in this analysis we
use all the ground state charmed hadrons: $\hc=\Do$, \Dp, \Ds\ and
\Lc, except for $\Xi_c^{0(-)}$ and $\Omega_c^0$ whose production rates
in \cc\ fragmentation are expected to be smaller than 1\% according to
the Lund model.  As two charmed hadrons are produced in
\cc\ fragmentation, the \jpcc\ cross section is given by half the sum
of the \jph\ cross sections. We extract \jph\ yields in both
\hc\ signal and sideband windows, using fits to \mel\ distributions
with signal and second order polynomial background functions.  The
\jp\ signal shape is obtained from MC simulation, with the small
difference in the \jp\ resolution between the MC and data
corrected. The \mel\ spectra are shown for \Do, \Dp, \Ds\ and
\Lc\ signal windows in Figs.~\ref{psiD} a), b), c) and d),
respectively; scaled sideband distributions are superimposed.  The
\jph\ yields are calculated as the difference between the \jp\ yields
in the signal window and the (scaled) sidebands. The fit results are
listed in Table~\ref{tab:psiD}.  We observe a significant excess
\jp\ signal in the \Do\ and \Dp\ signal windows with respect to the
corresponding sidebands, demonstrating large $\ee \to \jp \, \Do(\Dp)
\,X$ cross sections. An excess, with low significance, is also seen in
$\ee \to \jp \, \Ds(\Lc) \,X$.
% ratio of \Do, \Dp, \Ds\ and \Lc\ yields corrected for the efficiency
% is in agreement with the Lund predictions of the fractions of $\cc\to
% \hc X$, listed in Table~\ref{tab:psiD}.
\begin{figure}[tb]
\hspace*{-0.025\textwidth} \includegraphics[width=0.69\textwidth]
        {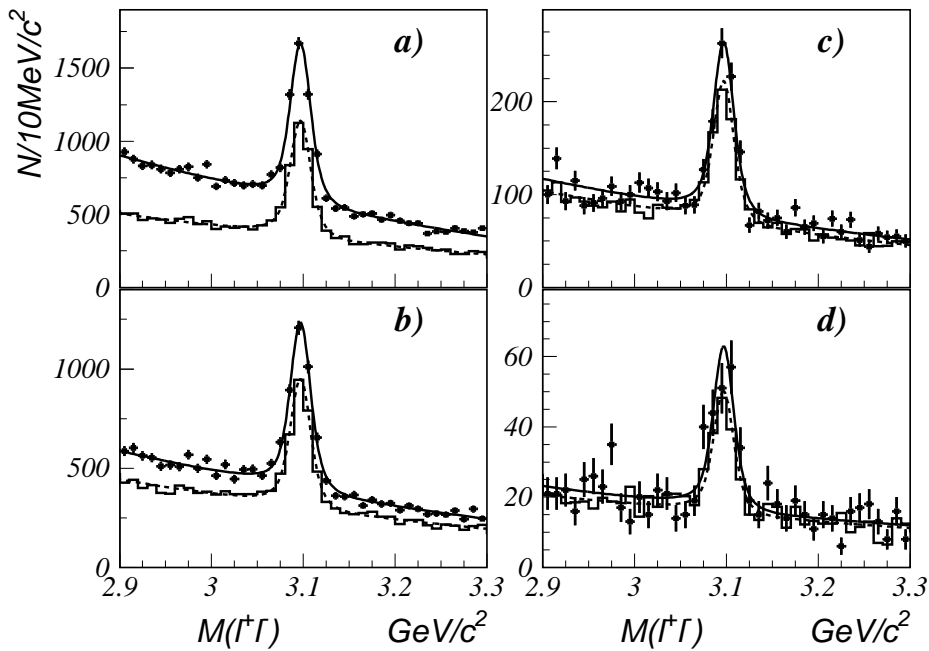}
\caption{\mel\ spectra for \hc\ signal (points with errors) and scaled
  \hc\ sideband windows (histograms), where \hc= a) \Do, b) \Dp, c)
  \Ds\ and d) \Lc. The curves represent the result of the fit; solid
  curves correspond to the \hc\ signal windows, and dashed curves to
  the \hc\ sidebands.}
\label{psiD}
\end{figure}
\begin{table}[tb]
\caption{\eejph\ signal yields and significances from 
  fits to the \mel\ spectra; the effective reconstruction efficiency
  times branching fraction is also shown.}
\label{tab:psiD}
\begin{center}
\begin{ruledtabular}
\begin{tabular}{l|cccc} 
& \Do & \Dp & \Ds & \Lc \\ \hline 

$N$ & $1072 \pm 108$ & \ $715 \pm 93$ \ & \ $129 \pm 42$ \ & \ $43 \pm
20$ \ \\

Significance \ & 10.1\,\sig & 7.8\,\sig & 3.2\,\sig & 2.2\,\sig \\

Efficiency$\,\times \, \BR$ \ & 0.041 & 0.047 & 0.022 & 0.020 \\
\end{tabular}
\end{ruledtabular}
\end{center}
\end{table}

Next, we measure the \jp\ momentum spectrum in inclusive production
and from the process \eejpcc. The inclusive \jp\ momentum spectrum is
obtained by fitting \el\ mass distributions in bins of \pjp\ with
signal and second order polynomial background functions. In the region
$\pjp\!< \!2.0\gevp$ only the continuum data is used; the \jp\ yields are
then scaled according to the ratio of luminosities. The ISR processes
$\ee \to \psi^{(\prime)}\, \gamma$ contribute to the selected sample
(with multiplicity greater than four) in the case of fake track
reconstruction and/or $\gamma$ conversion. This contribution is small
($\sim\!2\%$ of the total \jp\ rate), and is subtracted using a MC
simulation with \pp\ and \jp\ dielectron widths fixed to the PDG
values~\cite{pdg}. The final yield in each momentum bin, after
subtraction of QED background, is then corrected for the
\jp\ reconstruction efficiency and divided by the total luminosity.
The result, representing the differential cross section, is shown in
Fig.~\ref{mom}\,a) with open circles.

We calculate the momentum spectrum of \jp\ mesons from all double
charmonium processes, including \jp\ from cascade decays, and note
that the final state in $\ee \to \psi^{(\prime)} \, \cha$ events may
contain two \jp's. We use a MC simulation with the contributions of
double charmonium processes fixed to the results of the fit to data
(Fig.~\ref{cc2} and Table~\ref{tab:rmx}) to obtain this spectrum,
shown in Fig.~\ref{mom}\,a) with filled circles. The momentum spectrum
is peaked near the kinematical limit as expected for two-body
processes; ISR results in a tail to lower momentum values, and there
is an additional contribution at $\pjp\!\sim \! 3\gevp$ due to \jp's
from cascade decays.

To obtain the \jp\ momentum spectrum from the process \eejph, we
measure \jph\ yields in bins of \pjp. The fits to \mel\ spectra
(Fig.~\ref{psiD} and Table~\ref{tab:psiD}) are repeated in the
\hc\ signal and sideband windows for each bin, with the \jph\ yield
defined as the fitted \jp\ yield in the \hc\ mass window after
subtraction of the scaled yield in the \hc\ sidebands. Using the
continuum data it is possible to perform such fits below 2\gevp,
though with much larger statistical errors. The yield in each bin is
then corrected for the \jp\ and \hc\ reconstruction efficiencies,
using a MC simulation. The sum over all \hc\ weighted by a factor of
0.5 is plotted in Fig~\ref{mom}\,a) with filled squares and represents
the \jp\ momentum spectrum from the process \eejpcc, where the
\cc\ pair fragments into charmed hadrons. The sum of this distribution
and that from double charmonium production represents the
\jp\ momentum spectrum from the process \eejpcc; it is shown in
Fig.~\ref{mom}\,b) by the open squares. The difference between this
and the inclusive \jp\ spectrum is thus the spectrum from
\eejpncc\ events, where the system recoiling against the \jp\ is not
produced via a \cc\ pair (shown by the filled triangles in
Fig.~\ref{mom}\,b), to which the color-singlet \eejpgg\ and
color-octet \eejpg\ processes contribute.
\begin{figure}[htb]
\hspace*{-0.025\textwidth} \includegraphics[width=0.69\textwidth]
        {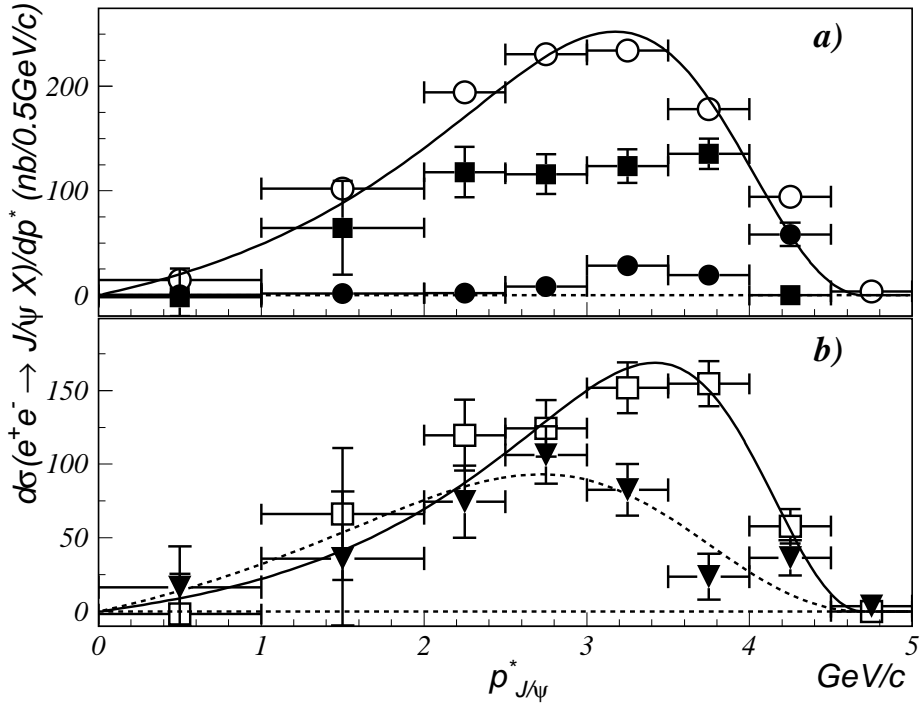}
\caption{\jp\ momentum spectra: a) inclusive (open circles), from
  \eejph\ (filled squares) and from double charmonium production
  (filled circles); b) the sum of all \eejpcc\ processes (open
  squares), from the \eejpncc\ processes (filled triangles). The results
  of fits to the Peterson function are shown in a) for the inclusive
  spectrum (solid curve); in b) for the processes \eejpcc\ (solid
  curve), and \eejpncc\ (dashed curve).}
\label{mom}
\end{figure}

The \jp\ efficiency corrected momentum spectra, shown in
Fig.~\ref{mom} for the processes \eejpx, \jpcc\ and \jpncc, are then
used to calculate the respective cross sections, after performing a
sum over all momentum bins. The results are presented in
Table~\ref{tab:sec}. The statistical errors are dominated by the
momentum interval $\pjp \!<\! 2.0 \gevp$, where only the small
continuum data sample is used.  To characterize the hardness of the
momentum spectrum, we perform fits using the Peterson
function~\cite{peter}; the parameters $\epsilon_{\rm{Peter}}$ for the
\eejpcc\ and \jpncc\ processes are listed in Table~\ref{tab:sec}.  For
completeness, the resulting cross sections $\sigma_{\rm{Peter}}$ are
also shown: they are consistent with the directly calculated values,
with statistical errors reduced by a factor of 1.5, as the fit
effectively extrapolates the high-momentum results into the
low-momentum region. Such results are model-dependent, and we rely
instead on the directly calculated values $\sigma$ for the cross
section.

\begin{table}[t]
\caption{Cross sections for the processes \eejpx, \jpcc\ and
  \jpncc\ ([pb]), and characteristics of the \jp\ spectra
  ($\epsilon_{\rm{Peter}}$, \ahel\ and \aprod); $\chi^2/n_{dof}$
  values for the corresponding fits are listed in parentheses.}
\label{tab:sec}
\begin{center}
\begin{ruledtabular}
\begin{tabular}
{l|ccc} 
& \jpx & \jpcc & \jpncc \\ \hline

$\sigma$ & \resx & \rescc & \resncc \\

$\sigma_{\rm{Peter}}$ & $1.19 \pm 0.01$ & $0.73 \pm 0.05$ & $0.48 \pm 0.07$
\\

$\epsilon_{\rm{Peter}}$ & \,$0.16 \pm 0.01$\,(8.9)\, & \,$0.10 \pm
0.02$\,(0.6) \,& \,$0.32_{-0.12}^{+0.16}$\,(1.6) \,\\ \hline 

\ahel & $\phantom{-}0.03\! \pm \!0.03$\,(0.6) &
$-0.19_{-0.22}^{+0.25}$\,(1.0) & $0.41_{-0.45}^{+0.60}$\,(1.2) \\

\aprod & $\phantom{-} 0.69 \! \pm \! 0.05$\,(3.3) &
$-0.26_{-0.22}^{+0.24}$\,(0.5) & $5.2_{-2.4}^{+6.1}$\,(0.3) 
\end{tabular}
\end{ruledtabular}
\end{center}
\end{table}

We note that unlike our first paper~\cite{behera} no correction for
the \nch\ requirement is applied for any of the process studied. For
\eejpncc\ such corrections are only possible by relying on a
model. However, for the process \eejpcc, the efficiency of the
$\nch\!>\!4$ requirement is more than 99\% if the \cc\ pair fragments
into charmed hadrons, as their decays lead to a large multiplicity in
the event. For double charmonium production the efficiency is $70\%$
according to the model used in the MC generator, and varies by $\pm
20\%$ with different charmonium decay models. As double charmonium
represents only $\sim\! 10\%$ of the total \eejpcc\ cross section, the
resulting correction is small, and included in the systematic error.

We also perform an angular analysis for the \eejpcc\ and
\eejpncc\ processes. This provides important information on the
production mechanisms, and allows the efficiency calculation to be
improved: the \jp\ reconstruction efficiency depends on both the
production angle (\tprod, the angle between the \jp\ momentum and the
beam axis in the CM frame) and the helicity angle (\thel, the angle
between the $\ell^+$ from \jp\ decay and the CM, seen from the
\jp\ rest frame). The MC simulation is adjusted to match the measured
distributions.

Angular distributions are obtained from fitted yields in bins of
\cprod\ and \chel, with an appropriate efficiency correction performed
bin-by-bin, for inclusive \jp, \jp\ from double charmonium production,
and \jp\ from \eejph.  The results are shown in Fig.~\ref{ang}.  The
inclusive \jp\ distributions (open circles) are obtained from
\jp\ yields.  Those for double charmonium production are obtained from
fits to the four $\RM((\cc)_{\tg})$ distributions, as for
Fig.~\ref{cc2} above.  Distributions for \eejph\ are obtained from
fitted \jp\ yields in appropriate \hc\ mass windows, after subtraction
of yields in the \hc\ sidebands. The distributions for \eejpcc\ (open
squares) are calculated as the sum of the corresponding distribution
for double charmonium production (with weight 1.0) and \eejph\ (with
weight 0.5). Distributions for the \eejpncc\ process (filled triangles)
are determined from the difference between \eejpx\ inclusive and
\jpcc\ distributions in each bin.

\begin{figure}[t]
\hspace*{-0.025\textwidth} \includegraphics[width=0.69\textwidth]
        {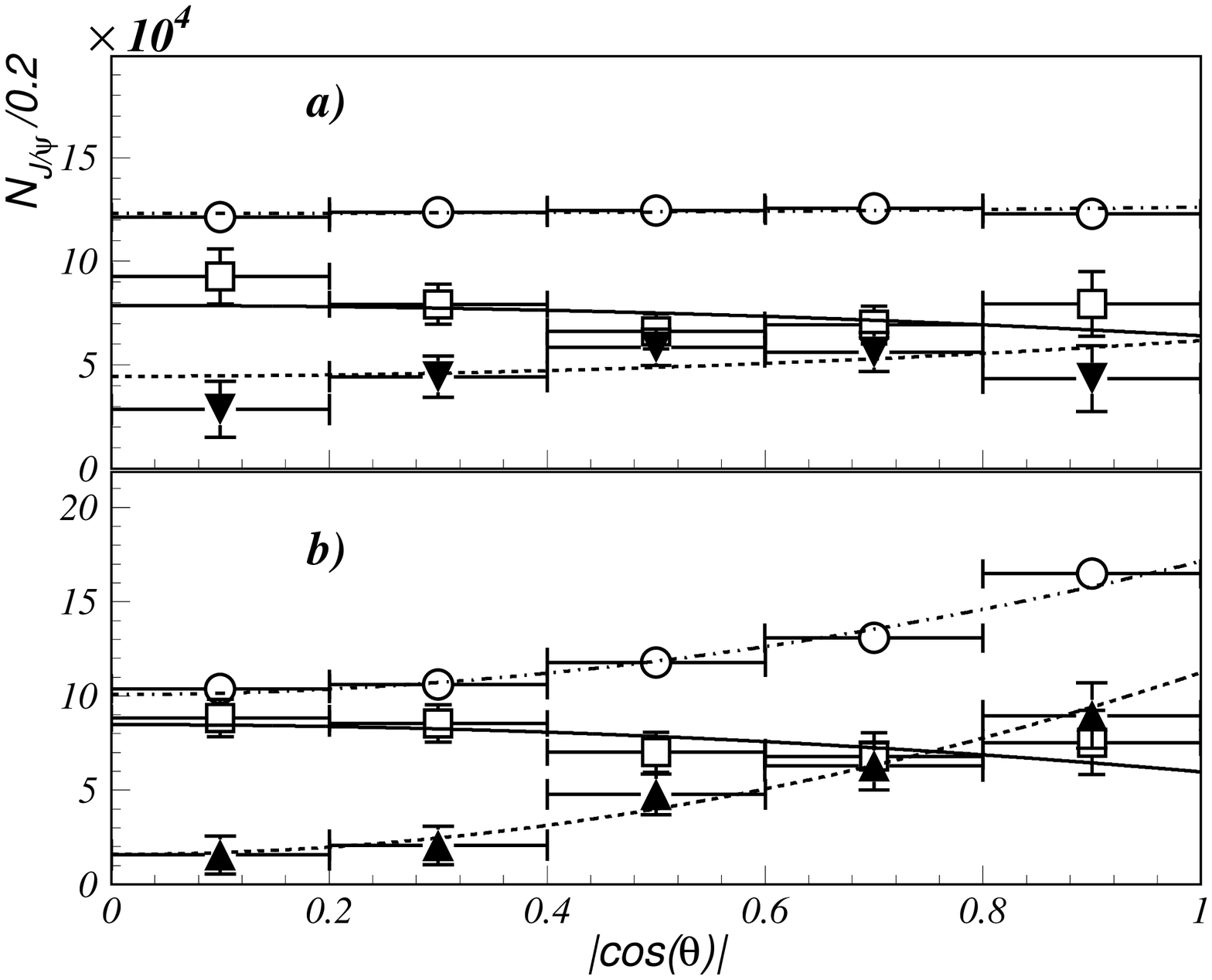}
\caption{Angular distributions (\chel\ in a), \cprod\ in b)) for
  inclusive \eejpx\ (open circles), \eejpcc\ (open squares), and
  \eejpncc\ processes (filled triangles).  The results of the fits
  described in the text are shown with the dash-dotted, solid, and
  dashed curves respectively.}
\label{ang}
\end{figure}

We fit the helicity angle distribution with a function $\sim\!(1+\ahel
\cos^2(\thel))$. While the production angle distributions are also
fitted with a function $\sim\!(1+\aprod \cos^2(\tprod))$, we note that
these distributions can differ from $1+\alpha \cos^2\theta$ due to ISR
or the contribution of the $\ee \to \gamma^* \gamma^* \to \jp \, X$
process to the \jpncc\ final state. The fits yield the parameters
\ahel\ and \aprod\ listed in Table~\ref{tab:sec}; the fit results are
shown in Fig.~\ref{ang}.

The systematic errors on the production cross sections for both
\eejpcc\ and \jpncc\ processes are summarized in
Table~\ref{tab:sys}. In the double charmonium production study,
systematic errors due to \jp\ yield fitting are determined as in our
previous papers~\cite{2cc, 2cc2}; we also perform variant fits
including final states with two charmonia with the same charge
conjugation eigenvalue.  In the study of associated production, we
consider changes in \jph\ yields under variation of the fitting
procedure (a 2-dimensional fit to ($M(\jp),M(\hc)$), a fit to the
$M(\jp)$ distribution in bins of $M(\hc)$, and to the $M(\hc)$ in bins
of $M(\jp)$), as well as variation of the signal and background
parameterizations, the fit ranges, and the binning.  The uncertainty
in \hc\ reconstruction efficiencies due to the unknown kinematics of
\cc\ fragmentation into charmed hadrons is small, due to the weak
dependence of reconstruction efficiency on \hc\ momentum, and is
included in the total systematic error.

When the integral \jp\ production and helicity angle distributions in
the MC simulation are tuned to those in the data, their correlations
are not taken into account. We assume the most conservative
correlations, resulting in the largest deviation of the
\jp\ reconstruction efficiencies that reproduce the integral
distributions. The resulting difference in efficiency is the largest
contribution to the systematic error. Other contributions come from
the uncertainty in the track and \ks\ (\La) reconstruction
efficiencies; from lepton, kaon and proton identification; and from
uncertainties in absolute \hc\ branching fractions.

\begin{table}[t]
\caption{Summary of the systematic errors on the cross sections shown,
	in percent.}
\label{tab:sys}
\begin{center}
\begin{ruledtabular}
\begin{tabular}{l|ccc}
Source & \jpx & \jpcc & \jpncc \\
\hline 

Fitting procedure      & $\pm 3$ & $\pm 5$ & $\pm 9$ \\

Selection              &  ---    & $\pm 3$ & $\pm 5$ \\

Angular distributions  & $\pm 4$ & $\pm 6$ & $\pm 10$ \\

\nch\ requirement      & --- & $^{+5}_{-0}$ & --- \\

ISR                    & ---     & $^{+4}_{-2}$ & $^{+4}_{-7}$ \\

Track reconstruction   & $\pm 2$ & $\pm 5$ & $\pm 8$ \\

Identification         & $\pm 2$ & $\pm 4$ & $\pm 7$ \\

$\mathcal{B}(\jp),~\mathcal{B}(\hc)$ & $\pm 1$ & $\pm 3$ & $\pm 3$ \\

\hline 

Total                 & $\pm 6$ & $^{+12}_{-11}$ & $\pm 20$ \\
\end{tabular}
\end{ruledtabular}
\end{center}
\end{table}

In summary, we have measured the cross sections for the processes
\eejpx, \jpcc\ and \jpncc\ to be $(\resx \sysx)\,$pb, $(\rescc
\syscc)\,$pb and $(\resncc \sysncc)\,$pb, respectively.  We therefore
conclude that \eejpcc\ is the dominant mechanism for \jp\ production
in \ee\ annihilation, contrary to the expectation from
NRQCD. Moreover, this cross section exceeds the perturbative QCD
prediction $\sigma(e^+e^-\to c\bar{c}c\bar{c})\approx\!0.3$\,pb
\cite{lik3}, which includes the case of fragmentation into four
charmed hadrons, rather than $\jp \cc$.  The cross section for \jpncc,
which can proceed via \eejpgg\ or $\jp\,g$, as well as $\ee \to \jp
\,\gamma^*$ diagrams, is of the same order as that for \jpcc. We have
measured the \jp\ momentum spectrum and the production and helicity
angle distributions from all three processes. For the
\eejpncc\ process, the \jp\ momentum spectrum is significantly softer
than that for \eejpcc, and the production angle distribution peaks
along the beam axis. We note that all the measured cross sections are
full (rather than Born) cross sections and include contributions from
cascade \jp, and that model-dependent corrections for the charged
track multiplicity requirement have not been performed.

We thank the KEKB group for excellent operation of the accelerator,
the KEK cryogenics group for efficient solenoid operations, and the
KEK computer group and the NII for valuable computing and Super-SINET
network support.  We acknowledge support from MEXT and JSPS (Japan);
ARC and DEST (Australia); NSFC and KIP of CAS (China); DST (India);
MOEHRD, KOSEF and KRF (Korea); KBN (Poland); MES and RFAAE (Russia);
ARRS (Slovenia); SNSF (Switzerland); NSC and MOE (Taiwan); and DOE
(USA).


\begin{thebibliography} {99}

\bibitem{nrqcd2a} P.~Cho and A.~K.~Leibovich, Phys. Rev. D {\bf 53},
  150 (1996); {\bf 53}, 6203 (1996); S.~Baek, P.~Ko, J.~Lee, and
  H.~S.~Song, J. Kor. Phys. Soc. {\bf 33}, 97 (1998), hep-ph/9804455.

\bibitem{nrqcd2d} F.~Yuan, C.-F.~Qiao, and K.-T.~Chao, Phys. Rev. D
  {\bf 56}, 321 (1997).

\bibitem{lik} V.~V.~Kiselev, A.~K.~Likhoded, and M.~V.~Shevlyagin,
  Phys. Lett. B {\bf 332}, 411 (1994).

\bibitem{lik2} A.V.~Berezhnoy, A.K.~Likhoded, Phys. Atom. Nucl. {\bf
  67}, 757 (2004), Yad.Fiz. {\bf 67}, 778 (2004).

\bibitem{2cc} K.~Abe, {\it {et al.}} (Belle Collab.),
  Phys. Rev. Lett. {\bf 89}, 142001 (2002).

\bibitem{kai} A.~B.~Kaidalov, JETP Lett. {\bf 77}, 349 (2003), Pisma
  Zh. Eksp. Teor. Fiz. {\bf 77}, 417 (2003).

\bibitem{Belle} A.~Abashian {\it et al.} (Belle Collab.),
  Nucl. Instr. and Meth. A {\bf 479}, 117 (2002); Z.~Natkaniec {\it et
    al.} (Belle Collab.), Nucl. Instr. and Meth. A {\bf 560}, 1
  (2006).

\bibitem{KEKB} S.~Kurokawa and E.~Kikutani, Nucl. Instr. and Meth. A
  {\bf 499}, 1 (2003), and other papers included in this Volume.

\bibitem{nim} E.~Nakano, Nucl. Instr. and Meth. A {\bf 494}, 402
  (2002).

\bibitem{cc} Charge-conjugate modes are included throughout this
  paper.

\bibitem{x4160} K.~Abe {\it et al.} (Belle Collab.),
  Phys. Rev. Lett. {\bf 98}, 082001 (2007).

\bibitem{y4260} C.Z.~Yuan {\it et al.} (Belle Collab.),
  Phys. Rev. Lett. {\bf 99}, 182004 (2007).

\bibitem{2cc2} K.~Abe {\it {et al.}} (Belle Collab.), Phys. Rev. D
  {\bf 70}, 071102 (2004).

\bibitem{pdg} C.~Amsler {\it et al.} (Particle Data Group),
  Phys. Lett. B {\bf 667}, 1 (2008).

\bibitem{lund} 
% T.~Sj\"ostrand, \textsc{Pythia} 5.7 / \textsc{Jetset}
%  7.4, CERN-TH.7112/93 (1993); 
T.~Sj\"ostrand, Comp. Phys. Commun. {\bf 82}, 74 (1994).

%\bibitem{pdg} W.~M.~Yao {\it et al.} (Particle Data Group)
%  J. Phys. {\bf G33}, 1 (2006).

\bibitem{peter} C.~Peterson {\it et al.}, Phys. Rev. D {\bf 27}, 105
  (1983).

\bibitem{behera} K.~Abe {\it et al.} (Belle Collab.),
  Phys. Rev. Lett. {\bf 88}, 052001 (2002).

%\bibitem{babar_2cc} B.~Aubert {\it {et al.}} (\bbr\ Collab.),
%  Phys. Rev. D {\bf 72}, 031101 (2005).

\bibitem{lik3} A.V.~Berezhnoy and A.K.~Likhoded,
  Phys. Atom. Nucl. {\bf 70}, 478 (2007).

\end{thebibliography}
\end{document}